\documentclass[12pt,a4paper]{article}
\usepackage[a4paper,left = 2.8cm, right = 2.8cm, top = 4cm, bottom = 5.5cm]{geometry}

\usepackage{mathptmx} 
\usepackage{avant}     

\usepackage[ruled,vlined]{algorithm2e}
\usepackage[affil-it]{authblk}
\usepackage[english]{babel}

\usepackage{latexsym,amsfonts,amssymb}
\usepackage{amssymb,amsmath,bm,mathrsfs,makeidx,amsfonts,graphicx,amsthm}
\usepackage[authoryear]{natbib}

\usepackage{epsfig}
\usepackage{setspace}
\usepackage{colordvi,multicol}
\usepackage{subfig}
\usepackage{graphicx}
\usepackage{graphics}
\usepackage{booktabs}
\usepackage{multirow}
\usepackage{lscape}
\usepackage{rotating}
\usepackage{multirow}
\usepackage{indentfirst}
\usepackage{caption}
\usepackage{array}
\usepackage{xcolor}

\usepackage{enumitem}
\usepackage{tabularx,booktabs}

\usepackage[colorlinks=true,citecolor=blue]{hyperref}
\newcolumntype{Y}{>{\centering\arraybackslash}X}

\newtheorem{thm}{Theorem}

\newtheorem{lem}{Lemma}

\newtheorem{remark}{Remark}

\numberwithin{equation}{section}
\newtheorem{pf}{Proof}
\def\ba{\begin{align*}}
\def\ea{\end{align*}}
\def\bao{\begin{align}}
\def\eao{\end{align}}
\def\begine{\begin{enumerate}}
\def\ende{\end{enumerate}}

\def\be{\begin{equation}}
\def\ee{\end{equation}}

\newcommand{\bbx}{{\boldsymbol x}}
\newcommand{\bbY}{{\boldsymbol Y}}
\newcommand{\bby}{{\boldsymbol y}}

\newcommand{\bbz}{{\boldsymbol z}}

\newcommand{\bbA}{{\boldsymbol A}}
\newcommand{\bba}{{\boldsymbol a}}
\newcommand{\bbB}{{\boldsymbol B}}

\newcommand{\bbD}{{\boldsymbol D}}

\newcommand{\bbb}{{\boldsymbol b}}

\newcommand{\bbI}{{\boldsymbol I}}

\newcommand{\bbQ}{{\boldsymbol Q}}

\newcommand{\bbR}{{\boldsymbol R}}

\newcommand{\bbr}{{\boldsymbol r}}

\newcommand{\bbU}{{\boldsymbol U}}

\newcommand{\bbu}{{\boldsymbol u}}

\newcommand{\bbone}{{\boldsymbol 1}}

\newcommand{\Sig}{\boldsymbol{\Sigma}}

\newcommand{\bbeta}{\boldsymbol{\beta}}

\newcommand{\bvar}{\boldsymbol{\varepsilon}}

\newcommand{\ex}{\mathbb{\boldsymbol{E}}}

\newcommand{\iu}{\mathrm{i}\mkern1mu}

\newcommand{\RN}[1]{%
  \textup{\uppercase\expandafter{\romannumeral#1}}%
}

\begin{document}

\def\spacingset#1{\renewcommand{\baselinestretch}%
{#1}\small\normalsize} \spacingset{1}

\title{Robust PCA for High Dimensional Data based on Characteristic Transformation} 
\author[1]{Lingyu He\thanks{Correspondence to: Dr. Lingyu He, College of Finance and Statistics, Hunan University, Changsha, Hunan 410000, China. Email: helingyu@hnu.edu.cn}}
\author[2]{Yanrong Yang}
\author[3]{Bo Zhang}
\affil[1]{Hunan University, China}
\affil[2]{The Australian National University, Australia}
\affil[3]{University of Science and Technology of China, China}

\maketitle
\date{}

\begin{abstract}
In this paper, we propose a novel robust Principal Component Analysis (PCA) for high-dimensional data in the presence of various heterogeneities, especially the heavy-tailedness and outliers. A transformation motivated by the characteristic function is constructed to improve the robustness of the classical PCA. Besides the typical outliers, the proposed method has the unique advantage of dealing with heavy-tail-distributed data, whose covariances could be nonexistent (positively infinite, for instance). The proposed approach is also a case of kernel principal component analysis (KPCA) method and adopts the robust and non-linear properties  via a bounded and non-linear kernel function. The merits of the new method are illustrated by some statistical properties including the upper bound of the excess error and the behaviors of the large eigenvalues under a spiked covariance model. In addition, we show the advantages of our method over the classical PCA by a variety of simulations. At last, we apply the new robust PCA to classify mice with different genotypes in a biological study based on their protein expression data and find that our method is more accurately on identifying abnormal mice comparing to the classical PCA.
\end{abstract}

\noindent%


\section{Introduction} 

Principal component analysis (PCA) (\citet{Anderson2003}, \citet{MR2036084}) is a widely used technique for data exploration and dimension reduction. As high-dimensional data are ubiquitously encountered with the fast development of modern technologies, PCA is drawing growing attention with its ability to summarize high-dimensional data by some low-dimensional projections (see, for example, \citet{donoho2000high}, \citet{johnstone2009statistical}, \citet{Yata2012}, \citet{lee2014convergence}, \citet{Shen2016SS}, \citet{morales2018asymptotics}, etc.). Mathematically the classical PCA is based on the eigendecomposition of the population covariance matrix, and the leading eigenvectors serve as the directions of the projections. The population covariance matrix, however, is very sensitive to the population distribution and the sample observations, for instance, the heavy-tailed distribution and outliers in data. Hence, the classical PCA has trouble in handling such data with bad qualities. With the explosion of the dimension, the heterogeneity, which is defined as the diversity of statistical properties of the data, becomes more and more common. For example, heavy-tailed variables are more likely to present along with the normal distributed features in high dimensional data and the classical PCA is sensitive to this kind of heterogeneity as shown in \citet{li1985projection}, \citet{He2020}. Other types of heterogeneity include heteroscedastic noise and outliers. In view of this, it is of great urgency to develop new dimension-reduction approach for the high-dimensional regime to deal with heterogeneities.

In this paper, we propose a new robust dimension-reduction approach for high-dimensio\\-nal data. In particular, the new method is especially useful for data drawing from heavy-tailed distributions, which is also considered in recent works including \citet{He2020}, \citet{Chen2021}, etc.
Imagine that the population distributions of the data have infinite second moments or even infinite first moments, then any dimension-reduction method depends on those moments, such as the classical PCA and the robust-covariance-based PCA (see \citet{croux2000principal} for example), are invalid. 
Motivated by this difficulty, we propose a novel robust PCA, in which the pivotal step is transforming the original data based on the form of the characteristic function. As discussed in \citet{Baxter1995} that transforming unusual-distributed data before analysis has special merits, our proposed method is robust to different styles of heterogeneity, especially to data with infinite population moments. The robustness mainly comes from the appealing properties of the transformation. Recall that for a real-valued random variable $y$, its characteristic function is $\phi(t) = \ex(exp\{\iu t y\})\ (t \in \mathbb{R},  \iu^2 = -1)$, which completely defines the probability distribution of $y$ and $|exp\{\iu t y\}| = 1$ for any $t$. Hence the transformation $z_i = exp\{\iu y_i\}\ (i  = 1, \ldots, p)$ retains the distribution information of $y_i\ (i  = 1, \ldots, p)$, and more noteworthily is a bounded random variable no matter $y_i\ (i  = 1, \ldots, p)$ is bounded or not. As a result, the proposed method is robust to heavy-tailed data by conducting the classical PCA on the transformed variables $z_i\ (i = 1, \ldots, p)$. 

The non-robustness issue of PCA has been studied in robust statistical analysis. A natural and simple idea is to replace the sample covariance matrix with a more robust estimator. \citet{croux2000principal} studied the influence functions and efficiencies of some robust covariance matrix estimators. Another approach, using a projection-pursuit index instead of the variance to measure the dispersion of the projections, is proposed by \citet{li1985projection}.  \citet{locantore1999robust} proposed robust PCA method by projecting the original data onto the unit sphere (centered at the spatial median), following which recent works such as \citet{Li2021} considered the properties of the covariance matrix for the transformed data in high dimensional regimes. In computer science, studying PCA in the view of a low-rank matrix approximation problem and minimizing the robust loss function has attracted attention (\citet{candes2011robust}). More recently, \citet{He2020} and \citet{Chen2021} studied the effect of heavy-tailedness using PCA-based approach in large-dimensional factor analysis. See, for example, \citet{RYS} and \citet{she2016robust} for more reviews. Our proposed method mainly contributes to the robust methods for heavy-tailed data without finite moments, similar to the data structure considered in \citet{He2020}. And we project the original data onto another space to achieve the robustness, which is similar to \citet{locantore1999robust} but with a different transformation. 

The proposed method is also a special case of Kernel PCA, which is widely used in pattern recognition and novelty detection, see \citet{DBLP:conf/nips/MikaSSMSR98}, \citet{HOFFMANN2007863}, and \citet{RYS} for example. The kernel PCA first maps the data into some feature space via a function and then performs PCA on the mapped data, which is the same procedure as our proposed method. The corresponding kernel function is given in Remark 2. \citet{DEBRUYNE20103007} showed that Kernel PCA with a bounded kernel is more robust than Kernel PCA with an unbounded kernel. This further supports the robustness of our method as our kernel function is bounded.
Moreover, as the kernel function is non-linear, the proposed method also helps explore the nonlinear relationship in the original data. 
Following the Kernel PCA literature \citet{GOL2007}, we study the properties of the reconstruction errors and the excess error (the difference between the optimal (population) reconstruction error and the empirical reconstruction error) of the PCA methods in Section \ref{robust:statpro01}. As Kernel PCA projects data into feature space and needs not have pre-images in the original space (\citet{DBLP:conf/nips/MikaSSMSR98}), the corresponding errors for our method refer to the transformed data. The upper bound of the excess error of the our method can be relatively small, even when the original data does not have finite variances.

Apart from the heterogeneity, the high-dimension itself is a crucial problem. Literatures including but not limited to \citet{johnstone2001distribution,LYB2011, Yata2012, lee2014convergence,Shen2016SS,wang2017asymptotics,cai2017limiting} have made effort to understand the behavior of empirical eigenvalues under different high-dimensional settings. For heavy-tailed data without finite variances, however, results from those literature are not applicable on the original data but still valid on the transformed data. On the other hand, it is well known from these literature that, the stronger spikeness of the population and the larger sample size allow larger dimension in consistently recovering the population eigenvalues from the empirical eigenvalues. We are interested in how does the transformation in our method affect the spike covariance structure if the original data is normal distributed and has a spiked covariance structure? We investigate via simulations the behavior of the largest $k$ eigenvalues for the original data and the transformed data under different settings. We find that the transformation retains the spiked structure under our simulated data. For the heavy-tail-distributed data, the empirical eigenvalues of classical PCA vary greatly, while the robust PCA gives more stable results.

In addition, reconstruction of the original data in our method is also important as that in Kernel PCA (\citet{DBLP:conf/nips/MikaSSMSR98}). We illustrate the advantage of our proposed method against the classical PCA in the sense of mean squared reconstruction error (MSE) with several examples. Those examples include data with heterogeneity in variances, data with outliers, and data from three different heave-tailed distributions. In total, we find that our proposed method can recover those data more accurately than the classical PCA. At last, we demonstrate an example of applying the method in real data analysis by analyzing the protein expression measurements of mice from \citet{higuera2015self}. Most of the proteins have heavy tails or extreme outliers in their expression levels, so it is essential to use robust methods on the data. The proposed method is used to classify mice with different genotypes based on their protein expression data. Comparing to the classical PCA, our proposed method can identify the mice with abnormal genotype more accurately.

The rest of this paper is organized as follows. Section \ref{robust:method} describes our proposed method in details. Section \ref{robust:statpro} studies the statistical properties. Simulations to illustrate the reconstruction performance under different cases are presented in Section \ref{robust:simulation}. Section \ref{robust:application} gives the example of a real data application.

\section{Methodology}
\label{robust:method}
Let us recall the settings of the classical Principal Component Analysis (PCA). Suppose we have $n$ data points $\bby_1, \ldots, \bby_n$, generated by a random vector $\bby = \left(y_1, y_2, \ldots, y_p\right)^{\top} \in \mathbb{R}^p$. The classical PCA aims to find a subspace $S \subset \mathbb{R}^p $ of dimension $k \ (k < p)$ that best fits those data points. Mathematically, the problem can be written as the following optimization problem:
\begin{equation*}
  \min_{\bbu, \bbU, \{\bbx_i\}} \sum_{i = 1}^n \|\bby_i - \bbu - \bbU\bbx_i\|^2 \quad\text{s.t.}\quad \bbU^{\top}\bbU = \bbI_k \quad \text{and}\quad \sum_{i = 1}^n\bbx_i = \mathbf{0},
\end{equation*}
where $\bbu$ is a point which represents the central of the subspace, $\bbU$ is a $p \times k$ matrix whose columns are the basis of the subspace and $\bbx_i \in \mathbb{R}^k$ is the vector of the new coordinates of $\bby_i$ in the subspace. The optimal solution to the classical PCA (Chapter 2.1.2 in \citet{RYS}) can be obtained as
\begin{equation*}
  \widehat{\bbu} = \frac{1}{n}\sum_{i = 1}^n \bby_i \quad \text{and} \quad \widehat{\bbx}_i = \widehat{\bbU}^{\top}\left(\bby_i - \widehat{\bbu}\right),
\end{equation*}
where $\widehat{\bbU}$ is a $p \times k$ matrix whose columns are the eigenvectors corresponding to the the largest $k$ eigenvalues of the sample covariance matrix
\begin{equation*}
  \widehat{\Sig}_n = \frac{1}{n}\sum_{i = 1}^n \left(\bby_i - \widehat{\bbu}\right)\left(\bby_i - \widehat{\bbu}\right)^{\top}.
\end{equation*}

Then $\widehat{\bbU}\widehat{\bbx}_i$ is the low-rank approximation of $\bby_i$ if we assume $\mathbb{E}(\bby) = \mathbf{0}$ and $\widehat{\bbu} = \mathbf{0}$ without lose of generality. However, it is well know that if the data contains extreme values or has a heavy-tailed distribution, the above optimization is not reliable and the solution $\widehat{\bbU}\widehat{\bbx}_i$ is not a good low-rank approximation to the original high-dimensional data. For example, consider the data points coming from a heavy-tailed distribution without a finite second moment, then the covariance matrix $\widehat{\Sig}_n$ will be extremely unreliable and invalid to make inferences on the population covariance matrix. 

To address this issue, we propose a new PCA method to obtain a good approximation of $\bby$,  which is robust to outliers and heavy-tailed distributions in this paper. The idea is to find a transformation which is robust to the heavy-tailed distribution or extreme values of the original data and then conduct the classical PCA on the transformed data instead. The details of the method is described as follows.

Let $\bbz = \left(z_1, z_2, \ldots, z_p\right)^{\top}$ be the transformed data of $\bby$, where the transformation is 
\begin{equation*}
	z_j = e^{\iu y_j}\ (j = 1,2, \ldots, p),
\end{equation*}
and $\iu$ is the imaginary unit. The reasons to make this transformation come from the special properties of $\bbz$. Firstly, $\bbz$ has finite second moments and contains most of the information in $\bby$ as it has the form of the characteristic function of $\bby$, which solves the problem that $\bby$ comes from heavy-tailed distributions, especially for those without the second moments. 
Secondly, the absolute value of $z_j$ equals $1$ for any $j = 1, \ldots, p$, which indicates that its variance is bounded. This property shrinks the effect of the possible outliers or extremely various variances on the result of the dimension reduction. Thirdly, due to the non-linear property of the transformation, it is capable of revealing non-linear relationship between components in $\bby$, which is different from the classical PCA which can only detect linear relationships. 

While there are desired properties with $\bbz$, it contains complex elements that make the situation complicated. On the other hand, according to Euler's formula, $z_j$ can be written as:
\begin{equation*}
z_j = e^{\iu y_j} = \cos{y_j} + \iu \sin{y_j}\quad (j = 1, \ldots, p).
\end{equation*}
Then if we define $\bbr = \begin{pmatrix}\cos{y_1},
       \ldots,
       \cos{y_p},
       \sin{y_1},
       \ldots,
       \sin{y_p}\end{pmatrix}^{\top}$, we have $\bbz$ as a linear transform of $\bbr$:
\begin{align*}
\bbz = \begin{pmatrix}
       1 & 0 & \cdots & 0 & \iu & 0 & \cdots & 0\\
       0 & 1 & \cdots & 0 & 0 & \iu & \cdots & 0 \\
       \vdots & \vdots & \ddots & \vdots & \vdots & \vdots & \ddots & \vdots \\
       0 & 0 & \cdots & 1 & 0 & 0 & \cdots & \iu
     \end{pmatrix}
     \begin{pmatrix}
       \cos{y_1}\\
       \vdots\\
       \cos{y_p}\\
       \sin{y_1}\\
       \vdots\\
       \sin{y_p}
     \end{pmatrix} := 
\begin{pmatrix}
       \bbI_p & \iu\bbI_p
     \end{pmatrix}
     \bbr.
\end{align*}
Assume there also exists a low rank subspace which best fits data points $\bbz_1, \ldots, \bbz_n$ generated from $\bbz$. Then, to find a good low-rank approximation of $\bbz_i (i = 1, \ldots, n)$, we only need conduct the classical PCA on $\bbr_i (i = 1, \ldots, n)$, which are real-valued random vectors.

Suppose $\Sig_{\bbr}$ is the covariance matrix of $\bbr$, and $\bbeta_1, \bbeta_2, \ldots, \bbeta_k$ are the orthonormal eigenvectors corresponding to the $k$ largest eigenvalues $\lambda_1 > \lambda_2 >\cdots > \lambda_k$ of $\Sig_{\bbr}$.  By the classical PCA method, $\bbr$ is approximated by
\begin{align*}
\widetilde{\bbr} &= \mathbb{E}(\bbr) + \sum_{j = 1}^k \bbeta_j \bbeta_j^{\top} \left(\bbr - \mathbb{E}(\bbr)\right) \\
&= \mathbb{E}(\bbr) + \left(\sum_{j = 1}^k \left(\bbr - \mathbb{E}(\bbr)\right)^{\top}  \bbeta_j \bbeta^{(\cos)\top}_j, \sum_{j = 1}^k \left(\bbr - \mathbb{E}(\bbr)\right)^{\top}  \bbeta_j \bbeta^{(\sin)\top}_j\right)^{\top},
\end{align*}
where $\bbeta^{(\cos)}_j = (\beta_{1,j}, \beta_{2,j}, \ldots, \beta_{p, j})^{\top}$ and $\bbeta^{(\sin)}_j = (\beta_{p+1,j}, \beta_{p+2,j}, \ldots, \beta_{2p, j})^{\top}$, which are the first half and the second half of $\bbeta_j$, respectively. 

Therefore, the low-dimensional approximation of $\bbz$ is
\begin{align*}
\widetilde{\bbz} &=  \begin{pmatrix}
       \bbI_p & \iu\bbI_p
     \end{pmatrix}
     \widetilde{\bbr}\\
     &= \begin{pmatrix}
       \bbI_p & \iu\bbI_p
     \end{pmatrix}
     \mathbb{E}(\bbr) + \\ \ &
     \begin{pmatrix}
       \bbI_p & \iu\bbI_p
     \end{pmatrix}
     \left(\sum_{j = 1}^k \left(\bbr - \mathbb{E}(\bbr)\right)^{\top}  \bbeta_j \bbeta^{(\cos)\top}_j,
     \sum_{j = 1}^k \left(\bbr - \mathbb{E}(\bbr)\right)^{\top}  \bbeta_j \bbeta^{(\sin)\top}_j\right)^{\top}
\end{align*}
\begin{align*}
    \quad = \mathbb{E}(\bbz) + \sum_{j = 1}^k  \bbeta^{(\cos)}_j \bbeta_j^{\top}  \left(\bbr - \mathbb{E}(\bbr)\right) + \iu \sum_{j = 1}^k  \bbeta^{(\sin)}_j \bbeta_j^{\top}  \left(\bbr - \mathbb{E}(\bbr)\right).
\end{align*}
With data points $\bbr_1, \ldots, \bbr_n$, it is straightforward to estimate $\mathbb{E}(\bbr)$, $\mathbb{E}(\bbz)$ and $\Sig_{\bbr}$ by 
\begin{equation*}
  \overline{\bbr} = \frac{1}{n}\sum_{i = 1}^n \bbr_i, \quad  \overline{\bbz} = \begin{pmatrix}
       \bbI_p & \iu\bbI_p
     \end{pmatrix}
     \overline{\bbr},  \quad \text{and} \quad
  \widehat{\Sig}_{\bbr, n} = \frac{1}{n}\sum_{i = 1}^n \left(\bbr_i - \overline{\bbr}\right)\left(\bbr_i - \overline{\bbr}\right)^{\top},
\end{equation*}
respectively. In addition, estimate $\bbeta_j \ (j = 1, \ldots k)$ by the eigenvectors $\widehat{\bbeta}_j \ (j = 1, \ldots k)$  of $\widehat{\Sig}_{\bbr, n}$. The method to estimate $k$ can be various and we use the accumulative variance as the criterion in the empirical analysis for simplicity. Other reasonable criterion can be applied under different purposes.

Finally, to recover the original data, we only need to transform back from $\widetilde{\bbz}_i (i = 1, \ldots, n)$. The approximation of $\bby_i (i = 1, \ldots, n)$ is:
\begin{align}
\widetilde{\bby}_i &= \frac{1}{\iu} \log (\widetilde{\bbz}_i) + 2h_i\pi\bbone \nonumber\\
&= \frac{1}{\iu} \log \left( \overline{\bbz} + \sum_{j = 1}^k  \widehat{\bbeta}^{(\cos)}_j \widehat{\bbeta}_j^{\top}  (\bbr_i - \overline{\bbr}) + \iu \sum_{j = 1}^k  \widehat{\bbeta}^{(\sin)}_j \widehat{\bbeta}_j^{\top}  (\bbr_i - \overline{\bbr})\right) +2h_i\pi\bbone, 
\label{eq1}
\end{align}
where $\log(\bba) = \left(\log(a_1), \log(a_2), \ldots, \log(a_n)\right)^{\top}${} for any $n$-dimensional vector $\bba$, and $h_i (i = 1, \ldots, n)$ is an integer which needs to be estimated in practice.

\begin{remark}
The computational algorithm is summarized in Algorithm \ref{AlgorithmR}.
\begin{algorithm}
\SetAlgoLined
\KwIn{Data $\bbY = [\bby_1, \dots, \bby_n] \in \mathbb{R}^{p \times n}$; Desired rank $\le p$.}
\KwOut{Low-dimensional representation of $\bbY$.}
\caption{{\bf Robust PCA for High Dimensional Data} \label{AlgorithmR}}
\textbf{Transformation Step}:\\
\nl Compute $\bbR = [\bbr_1, \dots, \bbr_n] \in \mathbb{R}^{(2p)\times n}$, where $\bbr_i = (\cos{y_{1,i}},
       \ldots,
       \cos{y_{p,i}},
       \sin{y_{1,i}},
       \ldots,
       \sin{y_{p,i}})^{\top}$\;
\textbf{PCA Step}:\\
\nl Compute the sample mean $\overline{\bbr} = n^{-1}\sum_{i = 1}^n \bbr_i$\;
\nl Compute the sample variance-covariance matrix $\widehat{\Sig}_{\bbr, n} = \frac{1}{n} \sum_{i=1}^{n} (\bbr_i - \overline{\bbr})( \bbr_i - \overline{\bbr})^{\top}$\;
\nl Conduct eigendecomposition on $\widehat{\Sig}_{\bbr, n}$ and get $\widehat{\bbeta}_1, \dots, \widehat{\bbeta}_{\widehat{k}}$, the eigenvectors corresponding to the largest $\widehat{k}$ eigenvalues of $\widehat{\Sig}_{\bbr, n}$\;
\textbf{Inverse Transformation Step}:\\
\nl Compute $\widehat{\bbeta}_j^{(\cos)} = (\beta_{1,j}, \dots, \beta_{p,j})^{\top}$ and $\widehat{\bbeta}_j^{(\sin)} = (\beta_{p+1,j}, \dots, \beta_{2p,j})^{\top}$\;
\nl Compute $\overline{\bbz} = \begin{pmatrix}
       \bbI_p & \iu\bbI_p
     \end{pmatrix}
     \overline{\bbr}$\;
\nl Compute $\widetilde{\bbz}_i = \overline{\bbz} + \sum_{j = 1}^{\widehat{k}} \widehat{\bbeta}_j^{(\cos)}\widehat{\bbeta}_j^{\top}(\bbr_i - \overline{\bbr}) + \iu \sum_{j = 1}^{\widehat{k}} \widehat{\bbeta}_j^{(\sin)}\widehat{\bbeta}_j^{\top}(\bbr_i - \overline{\bbr})$\;
\nl Compute $\widehat{h}_i = \arg\min_{h_i \in \mathbb{Z}}  \left|\bby_i - \left(\frac{1}{\iu} \log \widetilde{\bbz}_i + 2h_i\pi\bbone\right) \right|_2^2 \quad (i = 1, \ldots, n)$\;
\nl Compute $\widetilde{\bby}_i = \log(\widetilde{\bbz}_i)/\iu + 2\widehat{h}_i\pi\bbone$, $i = 1, \dots, n$.
\end{algorithm}
Note that $\widetilde{\bbz}_i (i = 1, \ldots, n)$ consists of complex numbers. The complex logarithm can have infinite many values, due to the periodicity of the complex exponential function. According to Euler's formula, those values are different by multiples of $2h\pi$. Therefore, in equation (\ref{eq1}), we need to find the $h_i$ to ensure that $\widetilde{\bby}_i$ is a good approximation to $\bby_i$. Hence, in practice, we estimate $h_i$ for each data point $\bby_i$ by
\begin{equation*}
\arg\min_{h_i \in \mathbb{Z}}  \left|\bby_i - \left(\frac{1}{\iu} \log \widetilde{\bbz}_i + 2h_i\pi\bbone\right) \right|_2^2 \quad (i = 1, \ldots, n).
\end{equation*} 
then \begin{align*}
	\widehat{h}_i =  \lfloor \left(\sum_{j = 1}^p\left(y_{ij} - \left(\log \widetilde{z}_{ij}\right)/\iu\right)\right)/(2\pi) +0.5 \rfloor, 
	\end{align*}
	where $\lfloor x \rfloor$ retains the integer part of $x$.
\end{remark}
\begin{remark}
    Our proposed method can be viewed as a special kind of nonlinear and Kernel PCA (Chapter 4.1 \citet{RYS}). The nonlinear transformation is $\phi(\cdot): \mathbb{R}^p \rightarrow \mathbb{R}^{2p}$, where
      $\phi(\bby) = (\cos{y_1}, \ldots, \cos{y_p}, \sin{y_1}, \ldots, \sin{y_p})^{\top}$
    and the kernel function is 
    \begin{align*}
      \kappa(\bby_i, \bby_j) = \phi(\bby_i)^{\top}\phi(\bby_j)
      = \sum_{m = 1}^p \cos{y}_{mi}\cos{y}_{mj} + \sum_{m = 1}^p \sin{y}_{mi}\sin{y}_{mj}.
    \end{align*}
    This shows that our method is capable to explore the non-liner relationship among the original data. In addition, we can compute the principal components with the kernel function described above according to the Kernel PCA algorithm (see Algorithm 4.1 in \citet{RYS} for example), which is particularly useful when the dimension $p$ is too large to compute the covariance matrix of the transformed data.
\end{remark}

\section{Statistical properties}
\label{robust:statpro}
We study the statistical properties of the robust PCA in this section. Firstly, we study inequilities of optimal and empirical reconstruction errors and give a general upper bound for the excess error of the PCA methods. It shows that the upper bound is not applicable for the classical PCA if the original data is heavy-tailed without finite variance. The robust PCA, however, can achieve a relatively smaller excess error due to that the proposed transformation ensures bounded variances for the transformed variables. We also discuss under which conditions the empirical reconstruction error of sample eigenvectors close to the optimal reconstruction error.
Secondly, we study the behavior of eigenvalues of both the original data and the transformed data.
By assuming a spike covariance structure on the original data, we find that the transformation in our proposed method still retain the spiked structure, which ensures that the proposed method can extract information from noises effectively. Furthermore, we show that the proposed robust PCA gives more stable empirical eigenvalues when the data is extremely heavy-tailed.

\subsection{The upper bound of the excess error}
\label{robust:statpro01}
In this section, we give the upper bound for aforementioned excess error and the order of the difference between the empirical reconstruction error of sample eigenvectors and the optimal reconstruction error in Theorem \ref{rethe1}. The Theorem \ref{rethe1} is generally hold for both the classical PCA and the newly proposed robust PCA. Through this theorem we show that, for extreme heavy-tailed data, the proposed method is able to achieve small excess error and empirical reconstruction error that closes to the optimal error. The classical PCA, however, may fail to do so under the same conditions.

Let us first introduce some notations and definitions in order to illustrate the results.
Suppose a random vector $\bby = \left(y_1, y_2, \ldots, y_p\right)^{\top} \in \mathbb{R}^p$ has mean $\mathbf{0}$ and covariance matrix $\Sig$. $\bby_1, \ldots, \bby_n$ are $n$ independent samples of $\bby$ and the corresponding sample covariance matrix is $\widehat{\Sig}$. Let $\bbeta_1, \ldots,\bbeta_p$ be the orthonormal eigenvectors corresponding to the eigenvalues of $\Sig$ in descending order, and $\widehat{\bbeta}_1, \ldots, \widehat{\bbeta}_p$ be those of $\widehat{\Sig}$. Denote $\bbB_k = (\bbeta_1, \ldots, \bbeta_k)$ and $\widehat{\bbB}_k = (\widehat{\bbeta}_1, \ldots, \widehat{\bbeta}_k)$ ($k < p$ is fixed). With $\bbB_k$, the basis of the optimal low-dimensional subspace, we have
\begin{align*}
\bby &= \sum_{i = 1}^k \bbeta_i\bbeta_i^{\top} \bby + \sum_{i = k+1}^p \bbeta_i\bbeta_i^{\top} \bby = \bbB_k\bbB_k^{\top}\bby  + \bbu(\bbB_k); \\
\bby_j &= \sum_{i = 1}^k \bbeta_i\bbeta_i^{\top} \bby_j + \sum_{i = k+1}^p \bbeta_i\bbeta_i^{\top} \bby_j =  \bbB_k\bbB_k^{\top}\bby_j + \bbu_j(\bbB_k) \quad (j = 1, \ldots, n).
\end{align*}
Similarly with $\widehat{\bbB}_k$, the empirical counterpart of $\bbB_k$, we have
\begin{align*}
\bby &= \sum_{i = 1}^k \widehat{\bbeta}_i\widehat{\bbeta}_i^{\top} \bby + \sum_{i = k+1}^p \widehat{\bbeta}_i\widehat{\bbeta}_i^{\top} \bby = \widehat{\bbB}_k\widehat{\bbB}_k^{\top}\bby + \bbu(\widehat{\bbB}_k);\\
\bby_j &= \sum_{i = 1}^k \widehat{\bbeta}_i\widehat{\bbeta}_i^{\top} \bby_j + \sum_{i = k+1}^p \widehat{\bbeta}_i\widehat{\bbeta}_i^{\top} \bby_j =  \widehat{\bbB}_k\widehat{\bbB}_k^{\top}\bby_j + \bbu_j(\widehat{\bbB}_k) \quad (j = 1, \ldots, n).
\end{align*} 
Therefore, the (true) reconstruction error with $\bbB_k$ and $\widehat{\bbB}_k$ can be written as
\begin{align*}
R(\bbB_k) &= \ex \left((\bby - \sum_{i = 1}^k \bbeta_i\bbeta_i^{\top} \bby)^{\top}(\bby - \sum_{i = 1}^k \bbeta_i\bbeta_i^{\top} \bby)\right) = \ex \left( \bbu(\bbB_k)^{\top}\bbu(\bbB_k) \right);\\
R(\widehat{\bbB}_k) &=  \ex \left( (\bby - \sum_{i = 1}^k \widehat{\bbeta}_i\widehat{\bbeta}_i^{\top} \bby)^{\top}(\bby - \sum_{i = 1}^k \widehat{\bbeta}_i\widehat{\bbeta}_i^{\top} \bby) \right) = \ex \left( \bbu(\widehat{\bbB}_k)^{\top}\bbu(\widehat{\bbB}_k) \right).
\end{align*}
We call $R(\bbB_k)$ the optimal error. The difference $R(\widehat{\bbB}_k) - R(\bbB_k)$ is the so-called excess error of $\widehat{\bbB}_k$ with respect to its optimal $\bbB_k$. Furthermore, the corresponding empirical reconstruction errors are
\begin{align*}
R_n(\bbB_k) &= \frac{1}{n} \sum_{j = 1}^n \left((\bby_j - \sum_{i = 1}^k \bbeta_i\bbeta_i^{\top} \bby_j)^{\top}(\bby_j - \sum_{i = 1}^k \bbeta_i\bbeta_i^{\top} \bby_j)\right) \\
&= \frac{1}{n} \sum_{j = 1}^n \left( \bbu_j(\bbB_k)^{\top}\bbu_j(\bbB_k) \right);\\
R_n(\widehat{\bbB}_k) &=  \frac{1}{n} \sum_{j = 1}^n \left( (\bby_j - \sum_{i = 1}^k \widehat{\bbeta}_i\widehat{\bbeta}_i^{\top} \bby_j)^{\top}(\bby_j - \sum_{i = 1}^k \widehat{\bbeta}_i\widehat{\bbeta}_i^{\top} \bby_j) \right) \\
&= \frac{1}{n} \sum_{j = 1}^n \left( \bbu_j(\widehat{\bbB}_k)^{\top}\bbu_j(\widehat{\bbB}_k) \right).
\end{align*}

Define
\begin{align*}
d_{k}:=\frac{\mathbb{E}\left(\bbu_j(\bbB_{k})^{\top}\bbu_j(\bbB_{k})\right)}{\mathbb{E}\left(\bby_j^{\top} \bby_j\right)}, \ \ \ k=1, 2, \ldots, \min\left(n, p\right).   
\end{align*} 

We have the following results for the true and empirical reconstruction errors:
\begin{thm}
\label{rethe1}
Suppose that $k$ is the number of spiked eigenvalues for the population covariance matrix, then we have 
\begin{align}
\label{the1.1}
P\left(\left|\left(R(\widehat{\bbB}_k) - R_n(\widehat{\bbB}_k)\right)\right| \le d_{k}\left(\sum^{p}_{i=1}\mathbb{E}\left(y_{ij}^2\right)\right)\sqrt{\frac{c\xi}{2n}} \right) \ge 1-2e^{-\xi},
\end{align}
and 
\begin{align}
\label{the1.2}
P\left(0 \le \left(R(\widehat{\bbB}_k) - R(\bbB_k)\right) \le 2d_{k}\left(\sum^{p}_{i=1}\mathbb{E}\left(y_{ij}^2\right)\right)\sqrt{\frac{c\xi}{2n}}\right) \ge 1 - 4e^{-\xi}, 
\end{align}
where $c$ is a constant number. 

In particular, as $\xi\rightarrow\infty$, if $d_{k}\left(\sum^{p}_{i=1}\mathbb{E}\left(y_{ij}^2\right)\right)\sqrt{\frac{c\xi}{2n}}\longrightarrow 0$, then we have
\begin{eqnarray}
\label{the1.3}
\left|\left(R(\bbB_k) - R_n(\widehat{\bbB}_k)\right)\right| 
=O_p\left(d_{k}\left(\sum^{p}_{i=1}\mathbb{E}\left(y_{ij}^2\right)\right)\sqrt{\frac{c\xi}{2n}}\right)=o_p(1). 
\end{eqnarray}
\end{thm}

\begin{remark}
We interpret the three differences of the reconstruction errors in Theorem~\ref{rethe1} as follows:
\begin{itemize}
	\item Inequality~(\ref{the1.1}) is the difference between the true reconstruction error and the empirical reconstruction error of the estimated low-dimensional subspace. It can be can be interpreted as a confidence interval on the true reconstruction error that can be computed from purely empirical data.
	\item  Inequality~(\ref{the1.2}) is the so-called excess error, which measures the distance between the estimated subspace and the optimal subspace in the sense of true reconstruction error.
	\item Finally, the difference in equation~(\ref{the1.2}) defines how far away the purely empirical reconstruction error to the optimal error.   
\end{itemize}
\end{remark}

In order to make the above upper bounds close to zero, we have two requirements:
\begin{itemize}[noitemsep]
  \item (1) $\xi$ is large enough, which ensures $e^{-\xi} \rightarrow 0$ and the probability close to 1;
  \item (2) $d_{k}\left(\sum^{p}_{i=1}\mathbb{E}\left(y_{ij}^2\right)\right)\sqrt{\frac{c\xi}{2n}} \rightarrow 0$ with ${k}$ being fixed and $p, n \rightarrow \infty$.
\end{itemize}
From the above conditions, we see that $\mathbb{E}(y_{ij}^2)$ is finite or not matters.
If $\mathbb{E}(y_{ij}^2)$ is infinite, it is hard to meet the second condition under high dimensional settings.
If $\mathbb{E}(y_{ij}^2)$ is finite, then $\left(\sum^{p}_{i=1}\mathbb{E}\left(y_{ij}^2\right)\right)\sqrt{c\xi/2n}$ $= O (p\sqrt{\xi/n})$ as $c$ is a constant. If $\xi$ is a large constant, for example $\xi = 10$, it is large enough to meet the first condition and the second condition becomes $d_k\frac{p}{\sqrt{n}} \rightarrow 0$ with ${k}$ being fixed and $p, n \rightarrow \infty$. Then to meet the second condition, we need $d_k = o(\sqrt{n}/p)$. For example, if we assume $p/\sqrt{n} = O(1)$, which is common in high-dimensional statistics, then we just require $d_k = o(1)$. It is worth mention that, $d_k$ should satisfy this condition well with a fixed $k$ if we assume a spiked covariance structure for the random vector. 

We have mentioned that Theorem~\ref{rethe1} holds for both the classical PCA and the proposed robust PCA. Similar statement can be find in \citet{GOL2007} where they give results that hold for both the classical PCA and the Kernel PCA.
To be more specific, the random variable $y_{ij}$ in Theorem~\ref{rethe1} refers to the original random vector for the classical PCA while to the transformed vector for the robust PCA. If the original data is heavy-tailed with infinite variances, the classical PCA is hard to meet the above conditions and may result in infinite excess error. On the other hand, the transformed data has finite variances even when the original data doesn't. Therefore, as long as the transformed data meets the aforementioned conditions for finite $\mathbb{E}(y_{ij}^2)$s, such as having a spiked covariance structure, the robust PCA will have empirical reconstruction error close to optimal error with order $o_p(1)$. 
When the original random vector has finite covariance matrix, so as the transformed vector, and there is not much difference between the classical PCA and the robust PCA in view of the order of the differences in Theorem~\ref{rethe1}. 
In summary, the proposed method is of great importance especially when the original data has infinite covariance matrix. 

The rest of this section is the proof of Theorem \ref{rethe1}. We will discuss more about the spiked covariance structure in next section (Section \ref{robust:statpro02}).

\begin{pf}
We make use of the following lemma of the concentration inequality to complete our proof.
\begin{lem}
\label{lemma1}
(McDiarmid(1989)) Let $X_1, \ldots, X_n$ be $n$ independent random variables taking values in $\mathcal{X}$ and let $Z = f(X_1, \ldots, X_n)$ where $f$ is such that
\begin{align*}
\sup\limits_{x_1,\ldots,x_n,x_i^{'} \in \mathcal{X}} |f(x_1, \ldots, x_n) - f(x_1, \ldots, x^{'}_i, \ldots, x_n)| \le c_i, \ \forall \ 1 \le i \le n,
\end{align*}
then
\begin{align*}
P[Z - \ex(Z) \ge \xi] \le e^{-2\xi^2/(c_1^2 + \ldots + c_n^2)} \quad \text{and} \quad
P[\ex(Z) - Z \ge \xi] \le e^{-2\xi^2/(c_1^2 + \ldots + c_n^2)}.
\end{align*}
\end{lem}

Let $\mathcal{X}$ be the set of all independent samples of $\bby$ and 
\begin{align*}
Z = f(\bby_1, \ldots, \bby_n) &= R(\widehat{\bbB}_k) - R_n(\widehat{\bbB}_k)
\\
&= \ex \left( \bbu(\widehat{\bbB}_k)^{\top}\bbu(\widehat{\bbB}_k) \right) - \frac{1}{n} \sum_{j = 1}^n \left( \bbu_j(\widehat{\bbB}_k)^{\top}\bbu_j(\widehat{\bbB}_k) \right).
\end{align*}
Then we have $\forall 1 \le i \le n$,
\begin{align*}
&\sup\limits_{\bby_1,\ldots,\bby_n,\bby_i^{'} \in \mathcal{X}} \left|f(\bby_1, \ldots, \bby_n) - f(\bby_1, \ldots, \bby^{'}_i, \ldots, \bby_n)\right| \\
&= \sup\limits_{\bby_1,\ldots,\bby_n,\bby_i^{'} \in \mathcal{X}} \left| \frac{1}{n}\left( \bbu_i(\widehat{\bbB}_k)^{\top}\bbu_i(\widehat{\bbB}_k)  - \bbu_i^{'}(\widehat{\bbB}_k)^{\top}\bbu_i^{'}(\widehat{\bbB}_k) \right)\right|.
\end{align*}
Thus in order to apply Lemma \ref{lemma1}, we only  need to find the upper bound of the above quantity.

The following evaluation
\begin{align*}
\mathbb{E}\left(\bbu_j(\bbB_k)^{\top}\bbu_j(\bbB_k)\right) 
=\sum^{p}_{i=1}\mathbb{E}\left(u_{ij}^2\right)
=O\left(d_k\sum^{p}_{i=1}\mathbb{E}\left(y_{ij}^2\right)\right),
\end{align*}
which indicates
\begin{eqnarray}\label{y1}
\bbu_j\left(\bbB_k\right)^{\top}\bbu_j\left(\bbB_k\right)=O_p\left(d_k\sum^{p}_{i=1}\mathbb{E}\left(y_{ij}^2\right)\right). 
\end{eqnarray}
Then from (\ref{y1}), we have for any $i$ and $j$,  
\begin{align*}
\left|\frac{1}{n}\left(\bbu_i(\bbB_k)^{\top}\bbu_i(\bbB_k) - \bbu_j(\bbB_k)^{\top}\bbu_j(\bbB_k)\right)\right| 
= O_p\left(\frac{d_{k}}{n}\sum_{i = 1}^p\mathbb{E}\left(y_{ij}^2\right)\right). 
\end{align*}
Let $c_j = \frac{{cd_{k}}}{n} \sum_{i = 1}^p\mathbb{E}\left(y_{ij}^2\right) \ (j = 1, \ldots, n)$ with $c$ being a constant which may be different from line to line. According to Lemma \ref{lemma1}, we have
\begin{align*}
P(\left|Z - \ex (Z)\right| \le t) &\ge 1 -  2e^{-2t^2/\left(\sum_{j = 1}^n\left(\frac{cd_{k}}{n} \sum_{i = 1}^p\mathbb{E}\left(y_{ij}^2\right)\right)^2\right)}\\
&=1-2e^{-2t^2/\left(\frac{cd_k^2}{n}\left[\sum^{p}_{i=1}\mathbb{E}\left(y_{ij}^2\right)\right]^2\right)}.
\end{align*}
Let $\xi = 2t^2/\left(\frac{cd_k^2}{n}\left[\sum^{p}_{i=1}\mathbb{E}\left(y_{ij}^2\right)\right]^2\right)$, which leads to $t = d_{k}\left(\sum^{p}_{i=1}\mathbb{E}\left(y_{ij}^2\right)\right)\sqrt{\frac{c\xi}{2n}}$. Then we can rewrite the above inequality as
\begin{align}
\label{e1}
P\left(\left|\left(R(\widehat{\bbB}_k) - R_n(\widehat{\bbB}_k)\right)\right| \le d_{k}\left(\sum^{p}_{i=1}\mathbb{E}\left(y_{ij}^2\right)\right)\sqrt{\frac{c\xi}{2n}}\right) \ge 1 - 2 e^{-\xi},
\end{align}
which is the first part of Theorem \ref{rethe1}.

For the second part of Theorem \ref{rethe1}, we first have
\begin{align}
\label{cond1}
R(\widehat{\bbB}_k) - R(\bbB_k) \ge 0 \quad \text{and} \quad
R_n(\widehat{\bbB}_k) - R_n(\bbB_k) \le 0
\end{align}
due to that $\bbB_k$ minimized the true reconstruction error and $\widehat{\bbB}_k$ minimized the empirical reconstruction error according to PCA. Hence we have
\begin{align*}
 0 &\le R(\widehat{\bbB}_k) - R(\bbB_k) \quad (\text{according to the first inequality in (\ref{cond1})})\\
&= \left(R(\widehat{\bbB}_k) - R_n(\widehat{\bbB}_k)\right) - \left(R(\bbB_k) - R_n(\bbB_k)\right) + \left(R_n(\widehat{\bbB}_k) - R_n(\bbB_k)\right) \\
&\le \left(R(\widehat{\bbB}_k) - R_n(\widehat{\bbB}_k)\right) - \left(R(\bbB_k) - R_n(\bbB_k)\right)\\
&\quad (\text{according to the second inequality in (\ref{cond1})})\\
&\le \left|\left(R(\widehat{\bbB}_k) - R_n(\widehat{\bbB}_k)\right)\right| + \left|\left(R(\bbB_k) - R_n(\bbB_k)\right)\right|.
\end{align*}
The first term is controlled by inequality (\ref{e1}). Following the same procedure, we also have
\begin{align*}
P\left(\left|R(\bbB_k) - R_n(\bbB_k)\right| \le d_{k}\left(\sum^{p}_{i=1}\mathbb{E}\left(y_{ij}^2\right)\right)\sqrt{\frac{c\xi}{2n}}\right) \ge 1 - 2 e^{-\xi}.
\end{align*}
Therefore, with probability $1 - 4e^{-\xi}$
\begin{align}
\label{e2}
P \left( 0 \le R(\widehat{\bbB}_k) - R(\bbB_k) \le  2d_{k}\left(\sum^{p}_{i=1}\mathbb{E}\left(y_{ij}^2\right)\right)\sqrt{\frac{c\xi}{2n}} \right) \ge 1 - 4e^{-\xi}.
\end{align}
Inequality (\ref{e1}) and (\ref{e2}) are the final results in Theorem \ref{rethe1}.
\end{pf}

\subsection{Behavior of the leading eigenvalues under spiked covariance structure}
\label{robust:statpro02}

As discussed before, how close the empirical reconstruction error to the optimal error depends highly on the covariance structure of data. In this section, we make assumptions on the covariance structure of the data and use some simulation to demonstrate behavior of leading eigenvalues under difference assumptions. We first assume that the original data is normal distributed and has a spiked covariance structure following assumptions in \citet{wang2017asymptotics} and \citet{cai2017limiting}. Under this assumption, we are interested in how does the transformation in our method affect the spike covariance structure? Next, we assume that the original data does not has finite covariance while the transformed data has a spiked covariance structure. We show under this assumption the estimations of the classical PCA vary greatly, while the robust PCA gives more stable results.  
\begin{remark}
	A variety of literature have made effort to understand the behavior of the empirical eigenvalues under the high dimensional settings. 
\citet{Yata2012,Shen2016SS} and others focused on the high-dimension, low-sample-size (HDLSS) case, where the dimension $p$ go to infinity with sample size $n$ fixed. Meanwhile,
\citet{johnstone2001distribution,LYB2011,lee2014convergence,wang2017asymptotics,cai2017limiting} considered situations that the sample size $n$ and the dimension $p$ both go to infinity. 
\citet{Shen2016JMLR} nicely characterized how the relationships of dimension, sample size and spike size affect PCA consistency. Particularly, we are interested in the spiked covariance model, of which the distribution of the empirical eigenvalues has been studied in \citet{wang2017asymptotics}, \citet{cai2017limiting}, and others. The spiked covariance model typically assumes that there are several eigenvalues larger than the rest. The larger eigenvalues are called the spiked eigenvalues, and the remaining ones are called the non-spiked eigenvalues. Specifically, \citet{wang2017asymptotics} and \citet{cai2017limiting} assume that the population covariance matrix has $k$ ($k/p \rightarrow 0$, p \ \text{is the number of dimension}) well separated spiked eigenvalues and the non-spiked eigenvalues are all bounded but otherwise arbitrary.
\end{remark}

In \citet{wang2017asymptotics}, the asymptotic normality of the spiked empirical eigenvalues was proved under a spiked covariance model (see Assumption 2.1 to 2.3 and Theorem 3.1 in \citet{wang2017asymptotics}).
Assume the population covariance model has $k$ spiked eigenvalues $\{\lambda_j\}_{j = 1}^k$, and the corresponding empirical eigenvalues are $\{\widehat{\lambda}_j\}_{j = 1}^k$. The theorem shows that $\widehat{\lambda}_j/\lambda_j (j = 1,2, \ldots, k)$ are asymptotic normal after some standardization and its bias is controlled by a term that contains rate $c_j = p/(n\lambda_j)$, where $n$ is the sample size and $p$ is the dimension. To make $\widehat{\lambda}_j$ asymptoticly unbiased, it requires $c_j \rightarrow 0$ for $j \le k$. If we assume the original data satisfies the above assumptions, the classical PCA should works well in estimating leading eigenvalues of the original data. Then how about our proposed robust PCA? Will the transformation still retains a spiked transformed covariance matrix, which ensures valid principle component results? We use a simulation from normal distributed data\footnote{We simulate $P \times 1$ ($P = 100$ in this case) vector $\bby_n\ (n = 1,2, \ldots, N)$ by 
\begin{align}
\bby_n = \sum_{i=1}^3 \alpha_i \bbb_i k_{i,n} + \bvar_{n},
\label{eqfsn}
\end{align}
where $\bbb_i\ (i = 1, 2, 3)$ are $P \times 1$ vectors generated by a QR decomposition, $k_{i, n}\ (i = 1,2,3, \ n =1,2, \ldots, N)$ are independently generated from standard normal $N(0, 1)$, and $\bvar_n \ (n = 1, 2, \ldots, N)$ are the $P \times 1$ error vectors with elements independently generated from $N(0, 1)$. Besides, $(\alpha_1, \alpha_2, \alpha_3) = (7, 5, 3)$. \\The details of the QR decomposition is as follows.
We intent to generate $\bbb_i\ (i = 1,2,3)$ such that $\bbb_1$, $\bbb_2$, and $\bbb_3$ are orthogonal to each other. We first generated a $P\times P$ matrix $\bbA$ with elements randomly from $N(0,1)$. Then decompose $\bbA$ into a product $\bbA = \bbQ\bbR$, where $\bbQ$ is an orthogonal matrix and $\bbR$ is an upper triangular matrix. We use the first three columns of the orthogonal matrix $\bbQ$ as the values of vectors $\bbb_i\ (i = 1,2,3)$ in the simulation studies.}  to illustrate the effect. We intend to demonstrate the changes of the population covariance structure after the proposed transformation, as well as the accurate of the empirical spiked eigenvalues for both the original and transformed data with different sample sizes.



\begin{figure}
\centering
\includegraphics[width=0.7\linewidth]{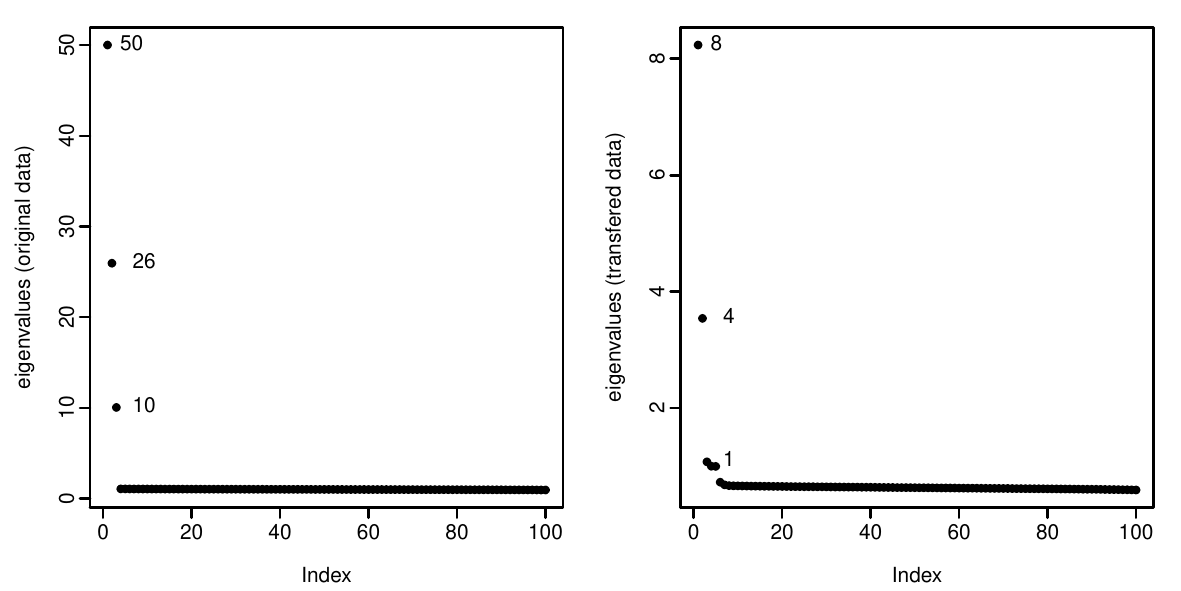}
\caption{Approximated population eigenvalues}
\label{fsn1}
\end{figure}

Let $\lambda_j^{cpca}$ and $\lambda_j^{rpca}$ be the two sets of population eigenvalues for original data and transformed data, respectively. Figure~\ref{fsn1} shows $\lambda_j^{cpca}$s and $\lambda_j^{rpca}$s (approximated) for the simulated example. The $\lambda_j^{cpca}$s and $\lambda_j^{rpca}$s are approximated by simulating the data with $n = 100000$ for $50$ times. We show $\lambda_j^{cpca}$s and $\lambda_j^{rpca}$s on the left and the right figures respectively. We see that both the original data and the transformed data have spiked covariance structure, although the number of leading eigenvalues may be different. That is to say, in this example, the transformation will retain a spiked structure when the original data has a spiked covariance structure. Hence, both the classical PCA and the proposed robust PCA should work well on this data.

Next we compare the estimations of the largest eigenvalue for the original data and the transformed data.
Note that  $\lambda_1^{rpca}$ is smaller than  $\lambda_1^{cpca}$. According to the rate $c_j = p/(n\lambda_j)$ we mentioned before, the transformation may affect the convergence rate of the eigenvalues. Table~\ref{spike1} present the effect empirically. We see from Table~\ref{spike1} that, $\widehat{\lambda}^{rpca}_1$ is more biased than $\widehat{\lambda}^{cpca}_1$ under the same sample size in this case. But as $n$ increasing, the bias of $\widehat{\lambda}^{rpca}_1$ tends to $0$ as desired. Moreover, in the high-dimensional cases ($n = 50, 100 \le p$), $\widehat{\lambda}^{rpca}_1$ varies less than $\widehat{\lambda}^{cpca}_1$.

\begin{table}[!htbp] \centering 
  \footnotesize
  \caption{Mean and SD of $\widehat{\lambda}^{cpca}_1/ \lambda^{cpca}_1 -1$ and $\widehat{\lambda}^{rpca}_1/ \lambda^{rpca}_1 -1$} 
  \label{spike1} 
\begin{tabularx}{\textwidth}{c *{12}{Y}}
\toprule
 & & \multicolumn{2}{c}{$50$}
  & \multicolumn{2}{c}{$100$}
  & \multicolumn{2}{c}{$500$}
  & \multicolumn{2}{c}{$1000$}
 & \multicolumn{2}{c}{$5000$}\\
\cmidrule(lr){3-4} \cmidrule(lr){5-6} \cmidrule(lr){7-8} \cmidrule(lr){9-10} \cmidrule(lr){11-12} 
 & & Bias & SD & Bias & SD & Bias & SD & Bias & SD & Bias & SD\\
\midrule
normal & cpca & $\bf{0.051}$ & $0.195$ & $\bf{0.030}$ & $0.137$ & $\bf{0.005}$ & $\bf{0.062}$ & $\bf{0.004}$ & $\bf{0.044}$ & $\bf{0.000}$ & $\bf{0.020}$ \\ 
 & rpca & $0.180$ & $\bf{0.127}$ & $0.094$ & $\bf{0.097}$ & $0.023$ & $0.070$ & $0.011$ & $0.064$ & $0.004$ & $0.057$ \\  
\bottomrule
\end{tabularx}
\end{table}

We have showed that both the classical PCA and the robust PCA work well in the normal case. In the case that the original data does not has finite covariance, however, the classical PCA will fail and the robust PCA should still work. We use a simulation to demonstrate the phenomenon. We simulate data the same as that in the normal case but with $k_{i,n}$ generated from t-distribution with degree of freedom $2$.  Note that in this case, the population eigenvalue of the transformed data $\lambda^{rpca}_1$ is well defined, but $\lambda^{cpca}_1$ does not exist.
Table~\ref{spike2} presents $\text{sd}(\widehat{\lambda}^{cpca}_1)/\text{mean}(\widehat{\lambda}^{cpca}_1)$ and $\text{sd}(\widehat{\lambda}^{rpca}_1)/\text{mean}(\widehat{\lambda}^{rpca}_1)$, which represent the variation of the largest empirical eigenvalues relative to their averages. It shows that $\widehat{\lambda}^{cpca}_1$ varies a lot while $\widehat{\lambda}^{rpca}_1$ is much more stable. In this extremely heavy-tailed case, the aforementioned asymptotic normal result for leading eigenvalues is valid on the transformed data but not applicable for the original data. Hence, we can't trust the results from classical PCA in this case. It provides a strong evidence that the classical PCA is not valid under extremely heavy-tailed data and our proposed robust PCA is necessary in such situations.

\begin{table}[!htbp] \centering 
  \footnotesize
  \caption{Variation of the largest empirical eigenvalues} 
  \label{spike2} 
\begin{tabularx}{\textwidth}{c *{7}{Y}}
\toprule
 & & 50 & 100 & 500 & 1000 & 5000 \\
\midrule
t(2) & cpca & $4.796$ & $16.082$ & $17.431$ & $4.752$ & $5.488$ \\ 
 & rpca & $0.105$ & $0.093$ & $0.074$ & $0.070$ & $0.068$ \\   
\bottomrule
\end{tabularx}
\end{table}

\section{Reconsturction performance under different situations}
\label{robust:simulation}
In this section, we illustrate the advantage of our proposed method (rpca) against the classic PCA (cpca) in recovery of orginal data under several scenarios. Throughout the simulations, we use mean squared error (MSE) of the approximation to measure the performance:
\begin{align*}
\text{MSE} = \sum_{n = 1}^{N}\|\widehat{\bby}_{n} - \bby_{n}\|_2^2/(NP), 
\end{align*}
where the $\widehat{\bby}_{n}$ is the approximation from rpca or cpca (both recovering at least 80\% of the total variance) and $\bby_n$ is the original data. $N$ is the sample size and $P$ is the dimension of $\bby_n$. The number of components $k$ is decided by threshold the cumulative sum of eigenvalues, that is
	\begin{align*}
	\widehat{k} = \min \{k: \sum_{i = 1}^k \widehat{\lambda}_i/\sum_{i = 1}^P \widehat{\lambda}_i \ge \gamma \}
	\end{align*}
	where $\widehat{\lambda}_i$ are the estimated eigenvalues and $\gamma$ is the threshold. We set $\gamma = 80\%$ in the following simulations. 

Example 1 shows the powerful ability of rpca to handle data with heterogeneity in variances. Example 2 demonstrates that rpca performs better than cpca when approximating data with outliers. Further, in Example 3, we simulate data from three different heave-tailed distributions, as well as the normal distribution as a benchmark, and we find that the rpca can recover those data more accurately than the cpca. 
Now let us discuss the simulations in details.  

\subsection{Example 1 : heterogeneity in variances}

The heterogeneity in variances is ubiquitous in real-life data, and the variables with extreme significant variances tend to dominate the results of classic PCA (\citet{MR2036084}). Hence, the information contained in other variables is masked, which makes the classic PCA less informative. In this example, we show that rpca can deal with this problem and recover the original data more precisely.

We simulate $\bby_n: P \times 1$ by $(\bby_n^{(1)\top}, \bby_n^{(2)\top})^{\top} (n = 1,2,\dots,N)$, where $\bby_n^{(1)}$ and $\bby_n^{(2)}$ are $(P/2) \times 1$ vectors generated by
\begin{align*}
\bby_n^{(1)} = \sum_{i=1}^3 \alpha_i \bbb_i^{(1)} k_{i,n}^{(1)} + \bvar_{n}^{(1)}, \quad \bby_n^{(2)} = \sum_{i=1}^3 \alpha_i \bbb_i^{(2)} k_{i,n}^{(2)} + \bvar_{n}^{(2)}
\end{align*}
where $\bbb_i^{(1)}$ and $\bbb_i^{(2)}\ (i = 1,2,3)$ are $(P/2) \times 1$ vectors independently generated by two QR decompositions. $k_{i, n}^{(1)} (i = 1,2,3, \ n =1,2, \dots, N)$, are independently generated from $N(0, 1)$ while $k_{i, n}^{(2)} (i = 1,2,3, \ n =1,2, \dots, N)$ are those from $N(0, 0.1)$. $\bvar_n^{(1)}$  and $\bvar_n^{(2)}$ are both the $(P/2) \times 1$ error vectors with elements independently generated from $N(0, 1)$. Besides, $(\alpha_1, \alpha_2, \alpha_3) = (7, 5, 3)$. 

\begin{figure}
\begin{minipage}[t]{0.48\textwidth}
\includegraphics[width=\linewidth]{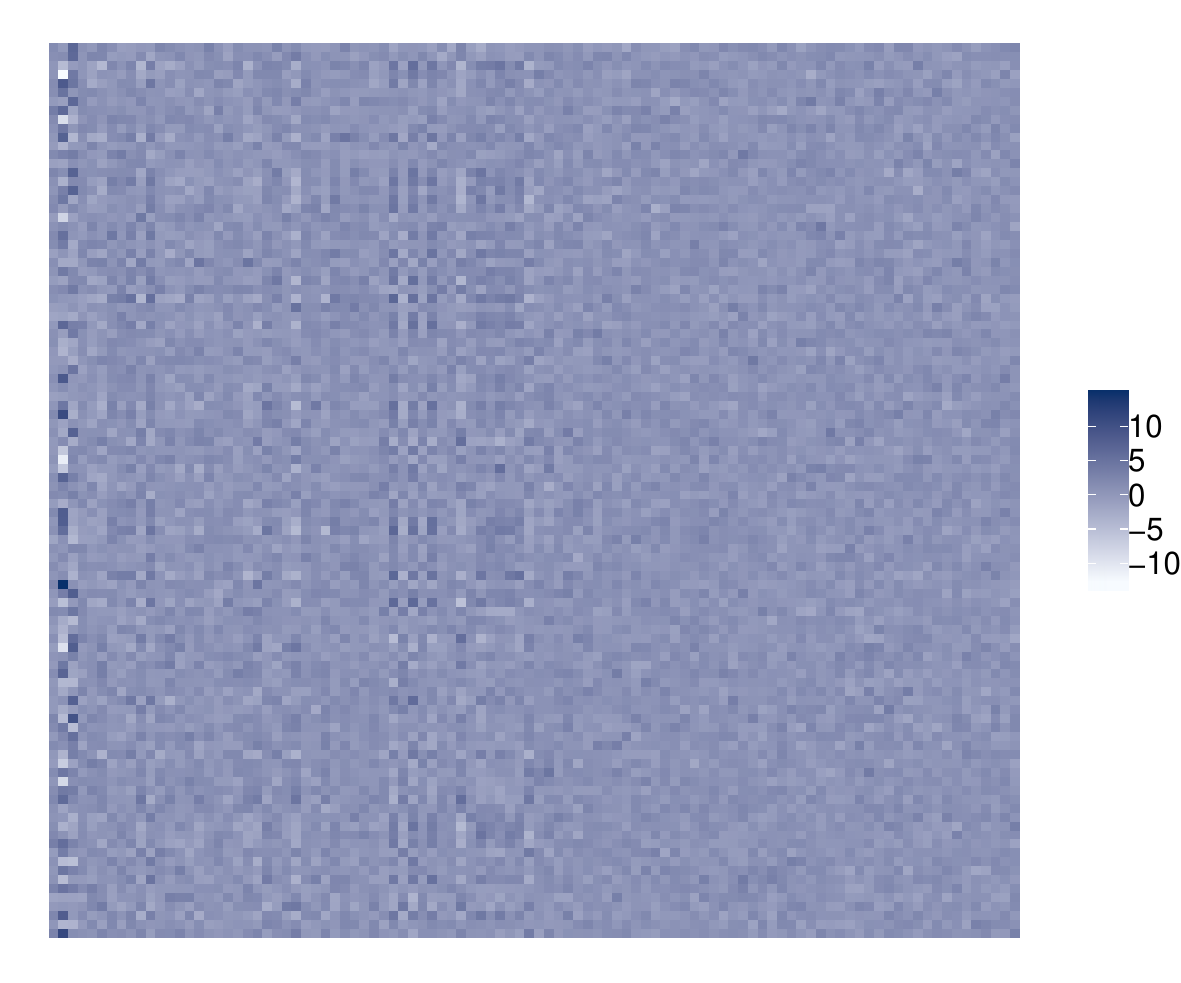}
\caption{example 1, data}
\label{RPCA.f11}
\end{minipage}
\hspace*{0.01cm}
\begin{minipage}[t]{0.48\textwidth}
\includegraphics[width=\linewidth]{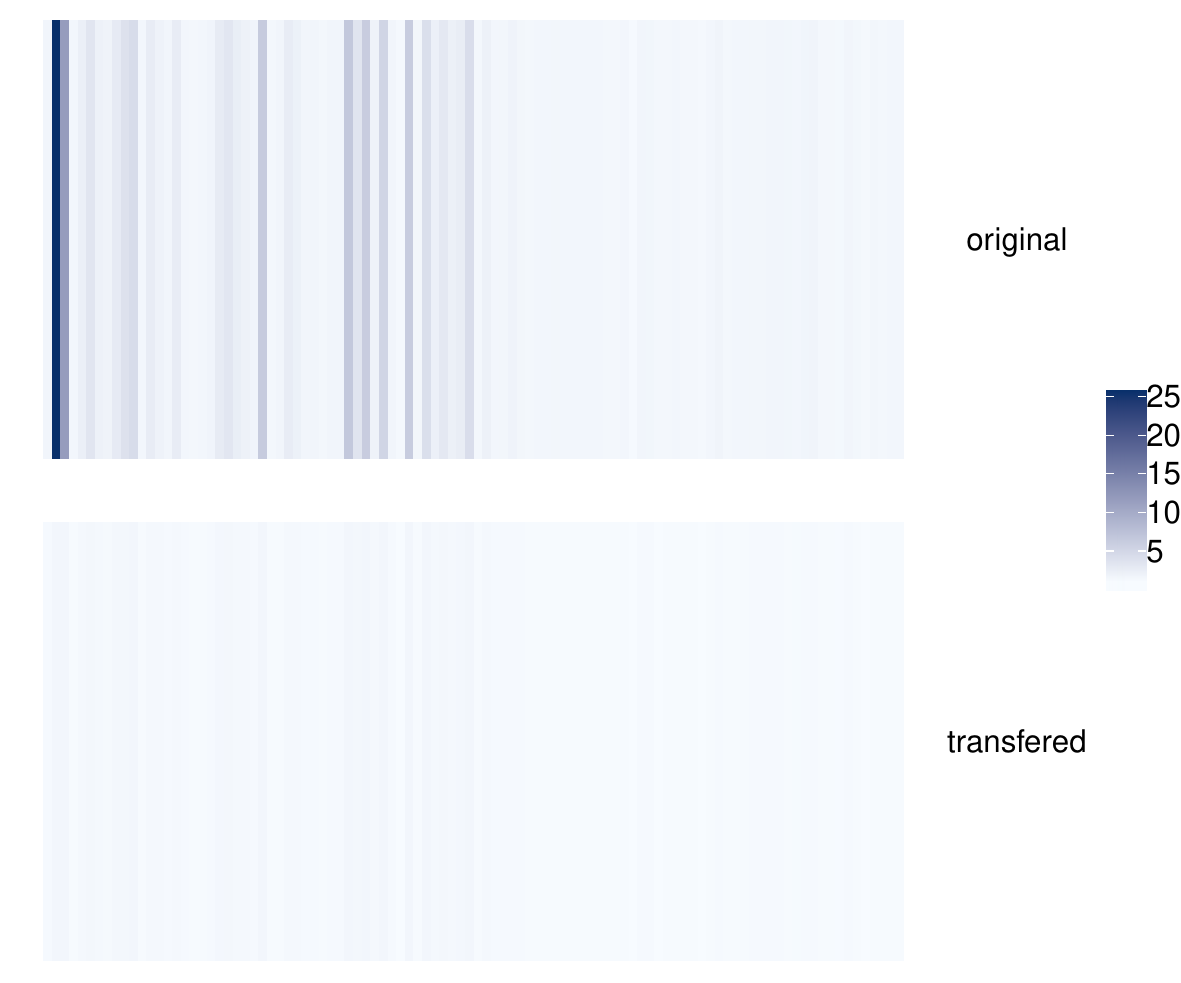}
\caption{example 1, variance}
\label{f12}
\end{minipage}
\end{figure}

Thus, $\bby_n$ consists of two parts with widely different variances. We can visualize the data and variance of a $100 \times 100$ sample matrix of $\bby_n$ in Figure \ref{RPCA.f11} and \ref{f12}. In both figures, the colour represents the size of the value: the darker the colour, the larger the value. Figure \ref{RPCA.f11} shows the original data matrix, and we can see clearly that some of the left parts have much more variations than the rest.
The top part of Figure \ref{f12}, which shows the sample variances of the original data, displays the widely differing variances more clearly. However, from the bottom part of Figure \ref{f12}, which shows the sample variances of $e^{\iu y_{in}} (i = 1,2, \dots, P)$, we see the differences in the variances are decreased after transforming the data. The transformation helps reduce the effect of the heterogeneity in variances on the results of PCA.

Next we compare the performance of cpca and rpca on approximating the data. We simulate this example for different sets of $(P, N):(50,40), (50, 100),$ $(100,100), (100, 200), (200, 190)$, which includes the situations of $P < N, P = N$ as well as $P > N$. Besides, although the value of $P$ and $N$ are not extremely large, we can consider $(50,40), (100,100)$ and $(200, 190)$ as high dimensional settings because the ratios $P/N \ge 1$. 
The average MSEs for 1000 simulations are shown in Table \ref{robust:t1}. It is clear that rpca performs better than cpca on recovering data with widely differing variances. For such data, classic PCA focus on those variables with large variances but ignores others which may be also very important. However, our proposed method automatically shrinks those differences, which is shown in Figure \ref{f12}, therefore results in a more accurate approximation. 
\begin{table}[!htbp] \centering 
  \footnotesize
  \caption{average MSE, 1000 simulations, Example 1} 
  \label{robust:t1} 
\begin{tabularx}{\textwidth}{c *{5}{Y}}
\toprule
 (P, N) & (50, 40) & (50, 100) & (100, 100) & (100, 200) & (200, 190) \\
\midrule
rpca & $\bf{0.203}$ & $\bf{0.211}$ & $\bf{0.201}$ & $\bf{0.204}$ & $\bf{0.193}$ \\ 
cpca & $0.498$ & $0.513$ & $0.357$ & $0.360$ & $0.279$ \\ 
\bottomrule
\end{tabularx}
\end{table}

\subsection{Example 2: outliers}

As the volume of data increasing, it is common to have outliers in the data. This example simulates data with outliers and shows that our proposed method is robust to such kind of data since the transformation can decrease the extreme of outliers. 

We first simulate $P \times 1$ vector $\bby_n$ by 
\begin{align*}
\bby_n = \sum_{i=1}^3 \alpha_i \bbb_i k_{i,n} + \bvar_{n}
\end{align*}
which is exactly the same as how we generated $\bby_n^{(1)}$ in Example 1 except with dimension $P$ instead of $P/2$. After simulating $N$ samples, we have a matrix $\bbY: P \times N$, whose columns consist of $\bby_1, \dots, \bby_N$.
Then we randomly replace $2.5\%, 6.4\%, 14.4\%$ of the elements in this matrix with values independently generated from $N(0, 6)$ or $N(0, 36)$. Thus about $2.5\%, 6.4\%$ or $14.4\%$ of the elements in $\bbY$ are outliers with variance $6$ or $36$. 

\begin{figure}
\begin{minipage}[t]{0.48\textwidth}
\includegraphics[width=\linewidth]{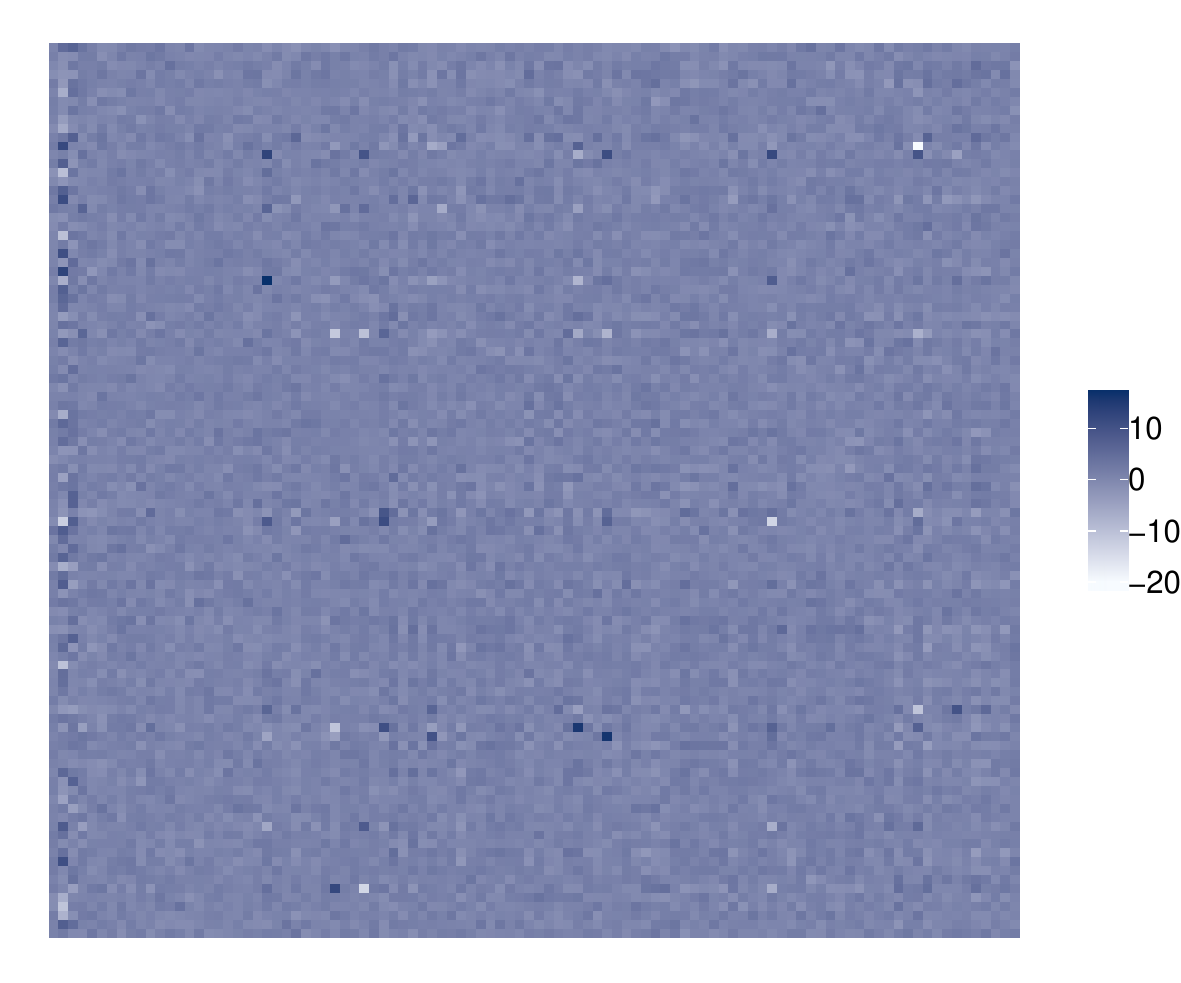}
\caption{example 2, data}
\label{f21}
\end{minipage}
\hspace*{0.01cm}
\begin{minipage}[t]{0.48\textwidth}
\includegraphics[width=\linewidth]{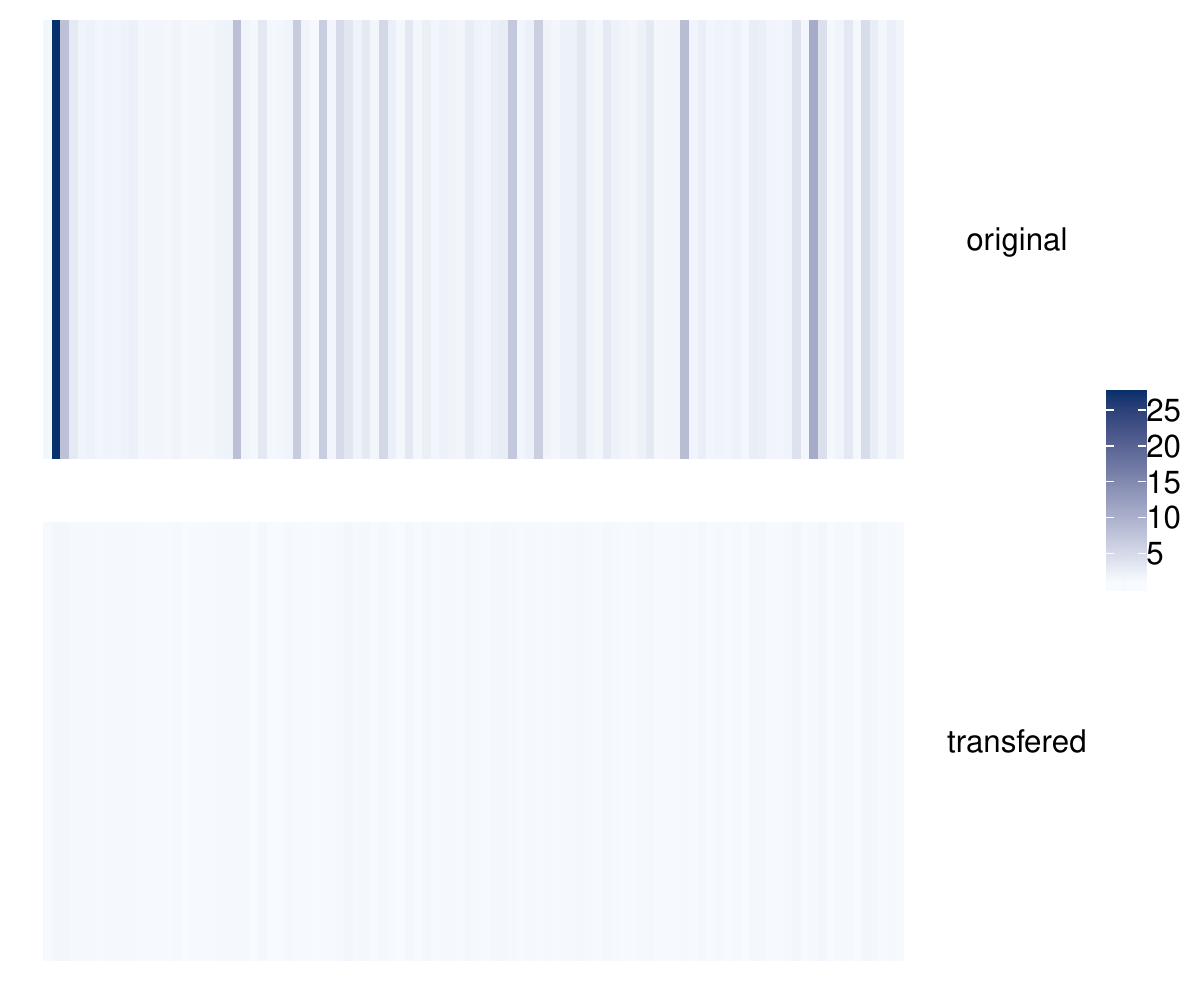}
\caption{example 2, variance}
\label{f22}
\end{minipage}
\end{figure}

The same as Example 1, we show values and variances of a $100 \times 100$ sample for Example 2 (outlier proportion is $14.4\%$ and from $N(0, 36)$) in Figure \ref{f21} and \ref{f22}. We can see clearly some squares with extremely darker or lighter colour than the others in Figure \ref{f21}, and those are outliers. From Figure \ref{f22}, we see there are some huge variances (top part of the figure) in the original data caused by the outliers, which is not a good sign for classical PCA, while our method can shrink those differences (bottom part of the figure) by the proposed transformation.
We try different sets of $(P, N)$ (which are the same as Example 1) and report the average MSEs of 1000 simulations in Table \ref{robust:t2}. We see that the rpca method performs robust to different settings of outliers. It is not surprising that rpca performs better than cpca, as rpca cuts back the differences between the average values and the outliers.

\begin{table}[!htbp] \centering 
  \footnotesize
  \caption{average MSE, 1000 simulations, Example 2} 
  \label{robust:t2} 
\begin{tabularx}{\textwidth}{c *{5}{Y}}
\toprule
 (P, N) & (50, 40) & (50, 100) & (100, 100) & (100, 200) & (200, 190) \\
\midrule

& \multicolumn{5}{c}{outliers: proportion $2.5\%$ from $N(0, 6)$} \\ 
\cmidrule(lr){2-6} 
rpca & $\bf{0.223}$ & $\bf{0.231}$ & $\bf{0.212}$ & $\bf{0.215}$ & $\bf{0.199}$ \\ 
cpca & $0.494$ & $0.513$ & $0.356$ & $0.360$ & $0.281$ \\
\midrule
& \multicolumn{5}{c}{outliers: proportion $6.4\%$ from $N(0, 6)$} \\ 
\cmidrule(lr){2-6} 
rpca & $\bf{0.224}$ & $\bf{0.233}$ & $\bf{0.213}$ & $\bf{0.216}$ & $\bf{0.199}$ \\ 
cpca & $0.496$ & $0.517$ & $0.359$ & $0.363$ & $0.284$ \\
\midrule
& \multicolumn{5}{c}{outliers: proportion $2.5\%$ from $N(0, 36)$} \\ 
\cmidrule(lr){2-6} 
rpca & $\bf{0.223}$ & $\bf{0.231}$ & $\bf{0.212}$ & $\bf{0.215}$ & $\bf{0.199}$ \\ 
cpca & $0.505$ & $0.524$ & $0.371$ & $0.375$ & $0.296$ \\
\midrule
& \multicolumn{5}{c}{outliers: proportion $14.4\%$ from $N(0, 36)$} \\ 
\cmidrule(lr){2-6} 
rpca & $\bf{0.225}$ & $\bf{0.234}$ & $\bf{0.215}$ & $\bf{0.218}$ & $\bf{0.201}$ \\ 
cpca & $0.586$ & $0.603$ & $0.451$ & $0.455$ & $0.377$ \\ 
\bottomrule
\end{tabularx}
\end{table}

\subsection{Example 3: heavy-tailed data}

Now we consider if rpca works well for data from different kinds of heavy-tailed distributions. There are a large amount of real-world data which have been proved to be heavy-tailed, therefore it is highly possible that a dataset with large dimensions contains heavy-tailed variables. We simulate data from t distribution, Pareto distribution and Cauchy distribution, which are all very common heavy-tailed distributions in real-world data. As a benchmark, we also simulate data from the normal distribution.

We simulate $P \times 1$ vector $\bby_n$ by 
\begin{align*}
\bby_n = \sum_{i=1}^3 \alpha_i \bbb_i k_{i,n} + \bvar_{n}
\end{align*}
which is the same as the first step in Example 2, except $k_{i, n} \ (i = 1,2,3, \ n =1,2, \dots, N)$, are independently generated from $N(0, 1)$ for the normal distribution, $t(2)$ for the t distribution, $pareto\ (scale = 0.5, shape = 1.5)$ for the Pareto distribution (by function `rpareto' in R package `VGAM'), and $cauchy\ (location = 0, scale = 1)$ for the Cauchy distribution. For this example, we try $(P,N) = (100, 100), (200,190)$ and the average MSEs of 1000 simulations are shown in Table \ref{t3}. 

Firstly, we see that on the normal-distributed data, the performance of rpca is better than that of cpca while the differences are not extremely large, which means on the normal-distributed data our proposed method is at least not worse than the classical PCA. Secondly, for the data from the three heavy-tailed distributions, rpca performs much better than cpca. One of the reasons for the worse performance of cpca is the uncertainty of the second moments of the heavy-tailed data. For example, the Cauchy distribution has no finite second moments, which makes the sample covariances estimated in cpca invalid and leads to the extremely bad performance shown in Table \ref{t3}.
However, the transformation of rpca guarantees that the transformed data has finite second moments, which ensures the feasibility of PCA on transformed data.

\begin{table}[!htbp] \centering 
  \footnotesize
  \caption{average MSE, 1000 simulations, Example 3} 
  \label{t3} 
\begin{tabularx}{\textwidth}{c *{8}{Y}}
\toprule
 (P, N) & \multicolumn{4}{c}{(100, 100)}
 & \multicolumn{4}{c}{(200, 190)}\\
\cmidrule(lr){2-5} \cmidrule(lr){6-9} 
 & Normal & t & Pareto & Cauchy & Normal & t & Pareto   & Cauchy \\
\midrule
rpca & $\bf{0.212}$ & $\bf{0.235}$ & $\bf{0.200}$ & $\bf{0.258}$ & $\bf{0.198}$ & $\bf{0.221}$ & $\bf{0.194}$ & $\bf{0.245}$ \\ 
cpca & $0.335$ & $1.174$ & $1.318$ & $127.242$ & $0.278$ & $0.840$ & $1.058$ & $388.595$ \\ 
\bottomrule
\end{tabularx}
\end{table}

\section{Empirical application}
\label{robust:application}
In this section, we performed the robust PCA on a real dataset to demonstrate an example of applying the method in real data analysis. The data, which has 77 variables and 1080 samples, comes from \citet{higuera2015self}, in which the details of the experiment and the measurements can be found. The data consists of the protein expression measurements of 77 proteins obtained from normal genotype control mice and Down syndrome (DS) mice, both with and without shock and drug treatments. There were 72 mice in the experiment, and 15 measurements of each protein per mouse were recorded. Thus there are 1080 (=72x15) expression measurements for each protein. We did a preprocessing step to deal with missing values.

\begin{figure}
\centering
\includegraphics[width=0.7\linewidth]{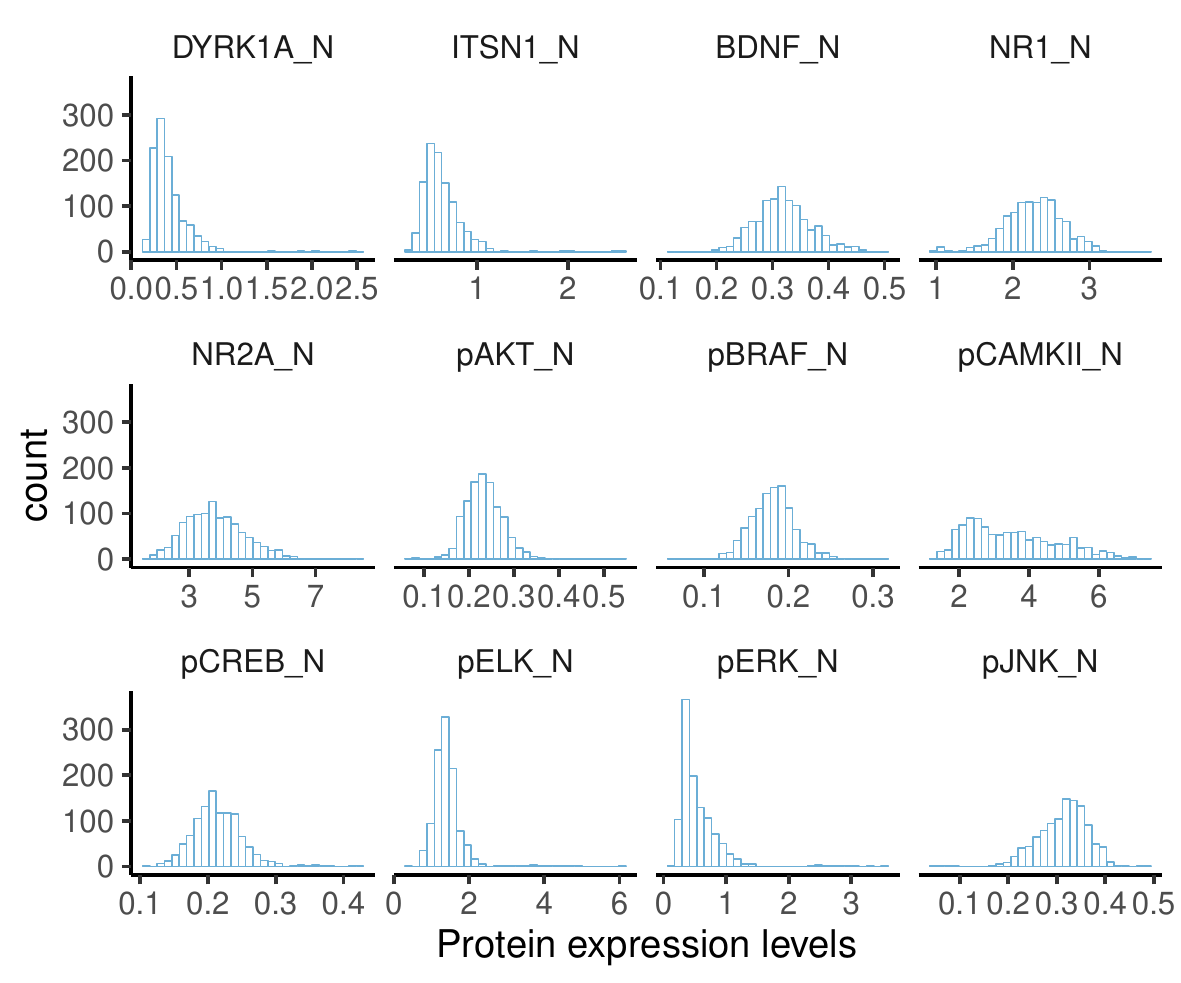}
\caption{The histogram of the expression measurements for the first 12 proteins}
\label{fa1}
\end{figure}

Figure \ref{fa1} shows the histograms of the expression measurements for the first 12 proteins in the data. We can see that although some proteins have nearly normal distributed expression levels, most of the proteins, such as DTRK1A, ITSN1, pCAMKII, and pERK, have heavy tails or extreme outliers in their expression levels. Thus, it is reasonable to statistically analyse this data with robust methods.

We first compare the approximations from robust PCA (rpca) and classic PCA (cpca) for the whole dataset under four criteria: the mean squared error (MSE) of the low rank representation, the number of principals we extracted based on threshold 0.8 of the total variance, the estimated smallest spiked eigenvalue, as well as the spiked ratio $P/(N\widehat{\lambda}_{\widehat{r}})$ (which we discussed in Section \ref{robust:statpro02}), which are shown in Table \ref{rwholeer}. Both of the spiked ratios are small, with the rpca one larger than that of cpca. It could indicate that rpca reduces the spiked eigenvalues and the smallest spiked one is more biased than that of cpca. On the other hand, with seven numbers of eigenvalues selected under threshold 0.8, rpca reaches a better approximation performance than cpca. It is worth mentioned that, although rpca is more flexible due to larger $\widehat{r}$ than cpca, the better out-of-sample performance provided later illustrates its appropriate flexibility.  

\begin{table}[!htbp] \centering 
  \footnotesize
  \caption{The comparison of rpca and cpca on the whole data} 
  \label{rwholeer} 
\begin{tabularx}{\textwidth}{c *{4}{Y}}
\toprule
   & MSE & $\widehat{r}$ & ratio ($\frac{P}{N\widehat{\lambda}_{\widehat{r}}}$) & $\widehat{\lambda}_{\widehat{r}}$ \\
\midrule
rpca & $\bf{0.009}$ & $7$ & $0.558$ & $0.126$ \\
cpca & $0.014$ & $3$ & $0.137$ & $0.512$ \\ 
\bottomrule
\end{tabularx}
\end{table}

One potential analysis for this dataset is using the protein expression levels to classify the mice. There were 38 control mice and 34 DS mice. The experiment in \citet{higuera2015self} involved shock and drug treatment for the treatment and control groups.
The shock treatment consisted of two types, one was context-shock (CS), which allowed the mice to explore a novel cage for several minutes and then gave a brief electric shock, and the other one was shock-context (SC), which did the inverse. Including the with and without the drug memantine, the mice are separated into eight groups. Hence, each group has 7 to 9 mice. Table \ref{nomic} shows the number of mice in each class. ``c'' represents the control group and ``t'' is the test group, which consists of DS mice. ``m'' represents the drug memantine and ``s'' is saline, which performs as a placebo.
\begin{table}[!htbp] \centering 
  \footnotesize
  \caption{Number of mice in each class, from \citet{higuera2015self}} 
  \label{nomic} 
\begin{tabularx}{\textwidth}{c *{2}{Y}}
\toprule
 & Classes & No. of mice \\ 
\midrule
Control mice & c-SC-s & $9$ \\
 & c-SC-m & $10$ \\ 
 & c-CS-s & $9$ \\ 
 & c-CS-m & $10$ \\  
\midrule
Down syndrome(DS) mice & t-SC-s & $9$ \\
 & t-SC-s & $9$ \\
 & t-SC-s & $7$ \\
 & t-SC-s & $9$ \\
\bottomrule
\end{tabularx}
\end{table}

We conduct a classification with a subset of the data for groups "c-CS-s" and "t-CS-s" by using the principal logistic regression. For these two groups, the shock and drug treatment were the same, but the genotype is different. One group consists of the normal mice while the other group consists of the DS mice. By \cite{higuera2015self}, the comparison of these two groups is biologically meaningful as it is related to the initial trisomy vs. control differences. We aim to use the protein expression levels through principal logistic regression to identify DS mice from the normal ones.

\begin{figure}
\centering
\includegraphics[width=0.6\linewidth]{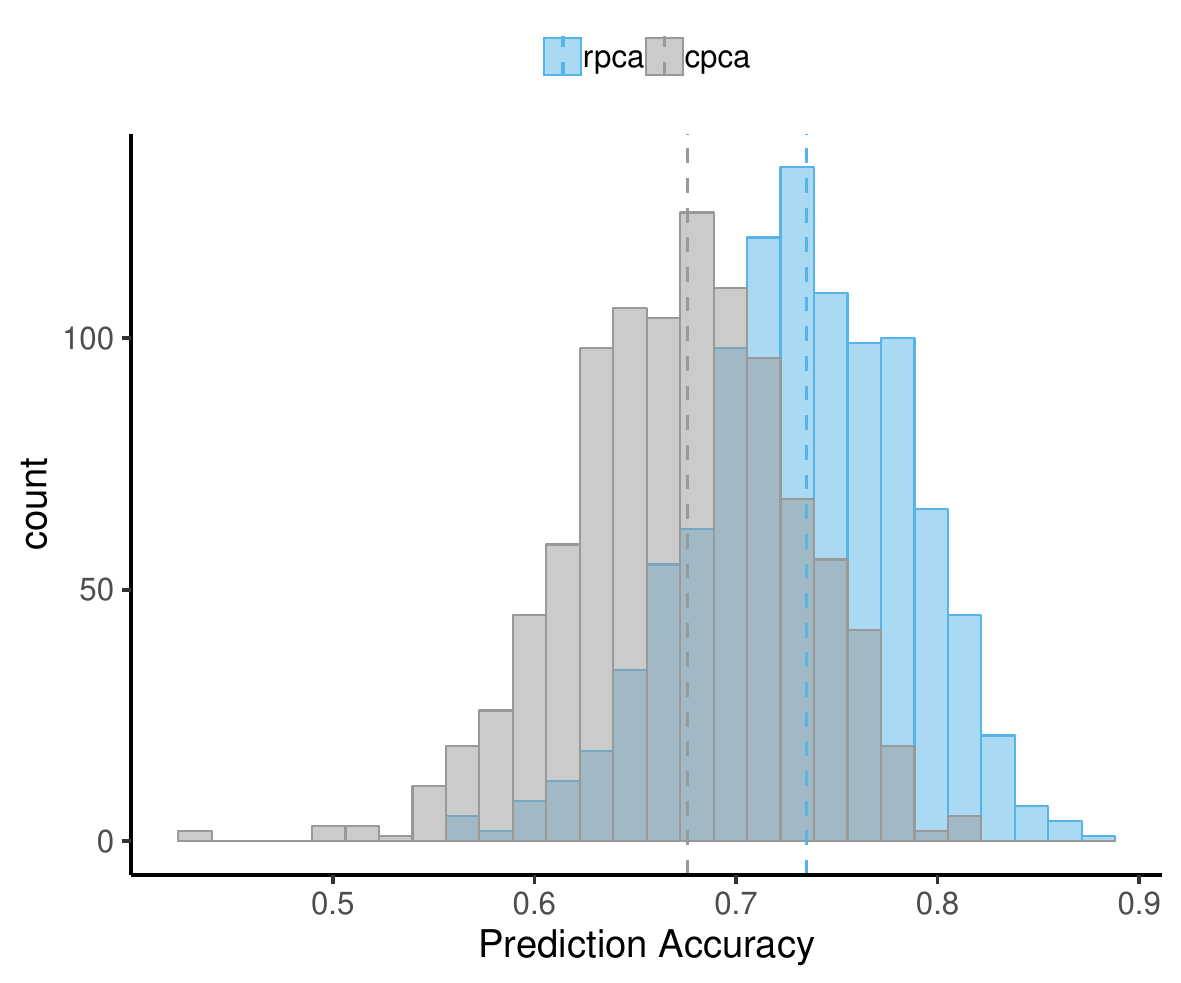}
\caption{Comparing the classification on mice data}
\label{fa2}
\end{figure}

The subset has 240 measurements and 77 proteins. 
We first split the data into training ($75\%$) and test sets ($25\%$) by random, in order to measure the prediction performance of the rpca and cpca by the cross-validation. Then we apply the rpca and cpca on the training data, extract the eigenvectors and construct the principal components as the design matrix for the logistic regression. For rpca, the principal design matrix is constructed by $\widehat{\bbB}^{\top}[\bbY^{\top}, \bbY^{\top}]^{\top}$, where $\widehat{\bbB}$ is the $(2P) \times \widehat{r}_1$ eigenvector matrix of the transformed training data, $\widehat{r}_1$ is the estimated number of eigenvalues of rpca, and $\bbY$ is the original training data with $P$ variables. For cpca, the corresponding principal design matrix is $\widehat{\bbD}^{\top}\bbY$, where $\widehat{\bbD}$ is the $P \times \widehat{r}_2$ eigenvector matrix of the original data, $\widehat{r}_2$ is the estimated number of eigenvalues of cpca, and $\bbY$ is the original data. Then, we use the principal design matrices as well as the class labels to fit logistic models and compute the prediction values for the test set. If the prediction value is larger than 0.5, we set it to be ``t-CS-s'', otherwise ``c-CS-s''. At last, we record the prediction accuracies for both of the methods. We repeat the process for 1000 times to ensure we have different training and test sets. Figure \ref{fa2} shows the histogram of the prediction accuracies and the mean accuracy for both methods. We can see that when using the principal design matrix constructed from robust PCA to fit the logistic model, almost all the prediction accuracy are larger than 0.5 and most of them are around 0.78. However, cpca performs much worse than rpca, with most of the prediction accuracy near 0.68. This is because heavy-tailed measurements and outliers affect the validity of the cpca, while the rpca method reduces those effects and results in a better performance.
Our proposed method can help identify DS mice from the normal ones by the protein expression levels effectively. This example shows that the robust PCA can definitely perform an essential role in classification models and also other statistic analysis.  

\section{Conclusion}
\label{rpca:conclusion}

In this paper, we addressed the challenge of applying the PCA on the high-dimensional data in the presence of various kinds of heterogeneities, especially the heavy-tailedness. Specifically, we proposed a robust PCA, based on a characteristic-function-type of transformation, to deal with the potential heterogeneities, which is particularly useful when the data is heavy-tailed (for example, with infinite variance). We show that the method is more robust than the classical PCA in the view of the excess error, assuming a spiked covariance structure for the data. We also studied the impact of the transformation on the spikeness of the spiked covariance structure. We illustrate with simulations that the transformation still keeps a well separable spiked covariance matrix. Particularly, the proposed method should work well when the original data has infinite variance, while the classical method is invalid. Simulations and empirical analysis show that the proposed robust PCA method is better than the classical PCA method, with the exist of heterogeneities in the data. As a by product, the proposed method is able to detect the non-linear relationships between the variables.

\bibliography{thesis}

\end{document}